\documentclass[aps,prl,reprint,showpacs,superscriptaddress]{revtex4-1}

\usepackage[T1]{fontenc}
\usepackage[utf8]{inputenc}
\usepackage{graphicx,color,rotating,pifont}
\usepackage{amsmath,amssymb}
\usepackage{mathtools}
\usepackage{siunitx}
\usepackage{bm}
\usepackage{ae}
\usepackage{dcolumn}
\usepackage{txfonts}
\usepackage{tensor}
\usepackage{braket}
\usepackage{booktabs}
\usepackage{xspace}

\usepackage{tikz}
\usetikzlibrary{positioning,calc,shapes,decorations.markings,decorations.pathmorphing}
\tikzset{asg/.cd,
  omega-vertex/.style={circle,solid,draw=black,fill=white,minimum size=5pt, inner sep=0pt},
  dbd-vertex/.style={coordinate},
  pline/.style={thick, postaction={decorate}, decoration={markings, mark=at position .5 with {\arrow[xshift=2pt]{stealth}}}},
  hline/.style={thick, postaction={decorate}, decoration={markings, mark=at position .5 with {\arrowreversed[xshift=-2pt]{stealth}}}},
  shift arrow/.style={/pgf/decoration/transform={xshift=#1}},
  shift arrow/.default=-2pt,
  dbd-2b/.style={decorate, decoration=snake},
  omega-2b/.style={densely dashed},
  neutron/.style={draw=blue},
  proton/.style={draw=red},
}

\usepackage{hyperref}

\DeclareMathSymbol{\NS}{\mathord}{AMSb}{"4E}

\DeclareSIUnit{\fm}{\femto\meter}

\definecolor{FGViolet}{rgb}{0.61,0.32,0.61}
\definecolor{FGDarkBlue}{rgb}{0,0,0.6}
\definecolor{FGBlue}{rgb}{0,0,0.8}
\definecolor{FGLightBlue}{rgb}{0.2, 0.6, 0.8}
\definecolor{FGGreen}{rgb}{0.2,0.7,0.2}
\definecolor{FGLightGreen}{rgb}{0.4,1,0.4}
\definecolor{FGYellow}{rgb}{1,0.95,0}
\definecolor{FGOrange}{rgb}{0.95,0.5,0.1}
\definecolor{FGRed}{rgb}{0.8,0,0}
\definecolor{FGWhite}{rgb}{1,1,1}
\definecolor{FGLightGray}{rgb}{0.8,0.8,0.8}
\definecolor{FGGray}{rgb}{0.5,0.5,0.5}
\definecolor{FGDarkGray}{rgb}{0.3,0.3,0.3}
\definecolor{FGBlack}{rgb}{0,0,0}

\begin{document}

\title{Advances in modeling nuclear matrix elements of  neutrinoless double beta decay}

\author{J. M. Yao}
\email{yaoj@frib.msu.edu}
 \affiliation{Facility for Rare Isotope Beams, Michigan State University, East Lansing, Michigan 48824-1321}

\date{\today}
   
\maketitle

\clearpage

 
  The neutrinoless double beta ($0\nu\beta\beta$) decay is a hypothetical weak process which manifests itself in the ``low-energy" environment of atomic nuclei  as two neutrons in a parent nucleus (A, Z) decays into two protons in a daughter nucleus (A, Z+2) with the emission of two electrons but no (anti)neutrinos \cite{Furry1939}.  This process violates lepton number -- an accidental global symmetry in the Standard Model. Its observation would  confirm lepton number violation in nature which has an important implication for the matter-antimatter asymmetry in the Universe. Besides, it would confirm the Majorana nature of neutrinos and provide the most promising way to determine their absolute mass scale. Therefore, the search for $0\nu\beta\beta$ decay has become a priority in nuclear and particle physics.

 The experimental search for the $0\nu\beta\beta$ decay is a great challenge as it is an extremely rare process if exists.  Currently, the best half-life lower limit (>$10^{25}$ years) is achieved in the experiments on $^{136}$Xe \cite{Gando2016},  $^{76}$Ge \cite{Agostini2018} and $^{130}$Te \cite{Adams2020}. 
  Next-generation experiments aiming to increase detector mass further to tonne-scale size and to reduce backgrounds as much as possible, are expected to reach a discovery potential of half-life $10^{28}$ years after a few years of running \cite{Dolinski2019}. 
  
    
 In the standard light-Majorana neutrino-exchange mechanism, the inverse of the decay  half-life  can be factorized as follows,
\begin{equation}
\label{half-life}
 [T^{0\nu\beta\beta}_{1/2}]^{-1} =  g^4_A G_{0\nu}
 \left(\dfrac{\langle m_{\beta\beta}\rangle}{m_e}\right)^2 
 \left\lvert M^{0\nu}\right\rvert^2
\end{equation}
where $g_A$ (unquenched value 1.27) is the nucleon axial charge, $m_e$ ($0.511$ MeV) is the electron mass. The nucleus-dependent phase-space factor $G_{0\nu}$ $(\sim10^{-14}{\rm yr}^{-1}$) can be evaluated precisely.  The effective Majorana neutrino mass $\langle m_{\beta\beta}\rangle=\bigl\lvert\sum^3_{k=1} U^2_{ek}m_k\bigr\rvert$  is a linear combination of neutrino masses $m_k$ weighted by the square of the elements $U_{ek}$ of the Pontecorvo-Maki-Nakagawa-Sakata (PMNS) matrix that mixes  neutrino flavors. 

 The null $0\nu\beta\beta$ decay signal from current experiments provides a constraint on the upper limits of effective neutrino mass $\langle m_{\beta\beta}\rangle$ if the decay is mediated by the exchange of light-Majorana neutrinos. In this scenario,  the next-generation tonne-scale experiments are expected to provide a definite answer on the mass hierarchy of neutrinos based on our current knowledge on the nuclear matrix element (NME) $M^{0\nu}$ of $0\nu\beta\beta$ decay.  The NME $M^{0\nu}=\langle \Psi_F\vert O^{0\nu}\vert \Psi_I\rangle$ cannot be measured, but must be determined from a theoretical calculation,  which is a challenge for nuclear theory. The ingredients: transition operator $ O^{0\nu}$ and nuclear wave functions $\vert\Psi_{I/F}\rangle$ for the initial and final nuclei, require a consistent treatment of both the strong and weak interactions and a precise modeling of nuclear many-body systems. The popularly used nuclear models based on either an energy density functional (EDF)  \cite{Rodriguez2010,Yao2015} universal for atomic nuclei throughout the nuclear chart or an effective interaction \cite{Menendez2009,Horoi2016} that is adjusted to atomic nuclei in a particular mass region  predict NMEs differing from each other by  a factor of about three, causing an uncertainty of an order of magnitude (or more) in the half-life for a given value of the neutrino mass. Resolving this discrepancy among model predictions has been one of the major tasks in nuclear theory community \cite{Engel2015}.  A great deal of effort has been devoted to understanding how the EDF or effective interaction and the model space affect the predicted NME \cite{Iwata2016,Jiao2017}. However, the systematic uncertainty turns out to be difficult to reduce because each model has its own phenomenology and uncontrolled approximations.

Thanks to the development in high-performance computing and the introduction of similarity renormalization group (SRG) into low-energy nuclear physics in the past decade, significant progress has been made in the {\em ab initio} modeling of atomic nuclei starting from nuclear interactions derived from chiral effective field theory (EFT) \cite{Weinberg1991}. In the chiral EFT, one can construct effective Lagrangians that consist of interactions that are consistent with the symmetries of quantum chromodynamics (QCD) and organized by an expansion in the power of $Q/\Lambda_\chi$. Here, $Q$ is a typical momentum of the interacting system, and $\Lambda_\chi$ is the breakdown scale of the chiral EFT, which is associated with physics that is not explicitly resolved. The power counting in the chiral EFT provides a convenient scheme to quantify the uncertainty from nuclear interactions and electroweak operators.  Starting from chiral interactions, {\em ab initio} methods are able to predict nuclear structure properties and single-beta decay rates of atomic nuclei up to mass number $A=100$ or even beyond \cite{Gysbers2019}. Nevertheless,  {\em ab initio} calculation of the NME of candidate $0\nu\beta\beta$ decay from first principles is not straightforward as the decay evolves deformed nuclei with strong collective correlations for which usual truncation schemes of nuclear many-body methods are ill suited.

Recently, we reported the first {\em ab initio} calculation for the NME of lightest candidate $0\nu\beta\beta$ decay from spherical $^{48}$Ca to deformed $^{48}$Ti with the in-medium generator-coordinate method (IM-GCM) starting from a particular SRG-softened chiral two-plus-three-nucleon interaction EM1.8/2.0 \cite{Yao2020}. This method provides a novel framework for describing arbitrary deformed medium-mass nuclei by combining the multi-reference in-medium similarity renormalization group (IMSRG) \cite{Hergert2016} with the GCM. The approach leverages the ability of the first method to capture dynamic correlations and the second to include collective correlations without violating symmetries. The $0\nu\beta\beta$ decay NME turns out to be $0.61^{+0.04}_{-0.05}$, which is in good agreement with the values from the most recent  two {\em ab initio} calculations with valence-space (VS) IMSRG approach \cite{Belley2020} and a leading-order approximation of coupled-cluster theory with  singles-doubles-and-triples excitations (CCSDT1) \cite{Novario2020} starting from the same nuclear interaction. In the VS-IMSRG calculation performed by Belley et al., the Hamiltonian, together with the $0\nu\beta\beta$ decay operator, is decoupled  into a given valence space using the single-reference IMSRG, which provides inputs for subsequent interacting shell model (ISM) calculation. Their predicted NME for $^{48}$Ca is 0.58(1).  In the CCSDT1 calculation carried out by Novario et al.,  equation-of-motion techniques are used to compute the NME, in which the ground state of one nucleus is represented as an isospin excitation of another nucleus in the decay. They predicted the NME in between $0.25$ and $0.75$, where the upper and lower boundary values are obtained from the choice of initial and final state as reference state, respectively.

\begin{figure}[]
\centering
\includegraphics[width=8.7cm]{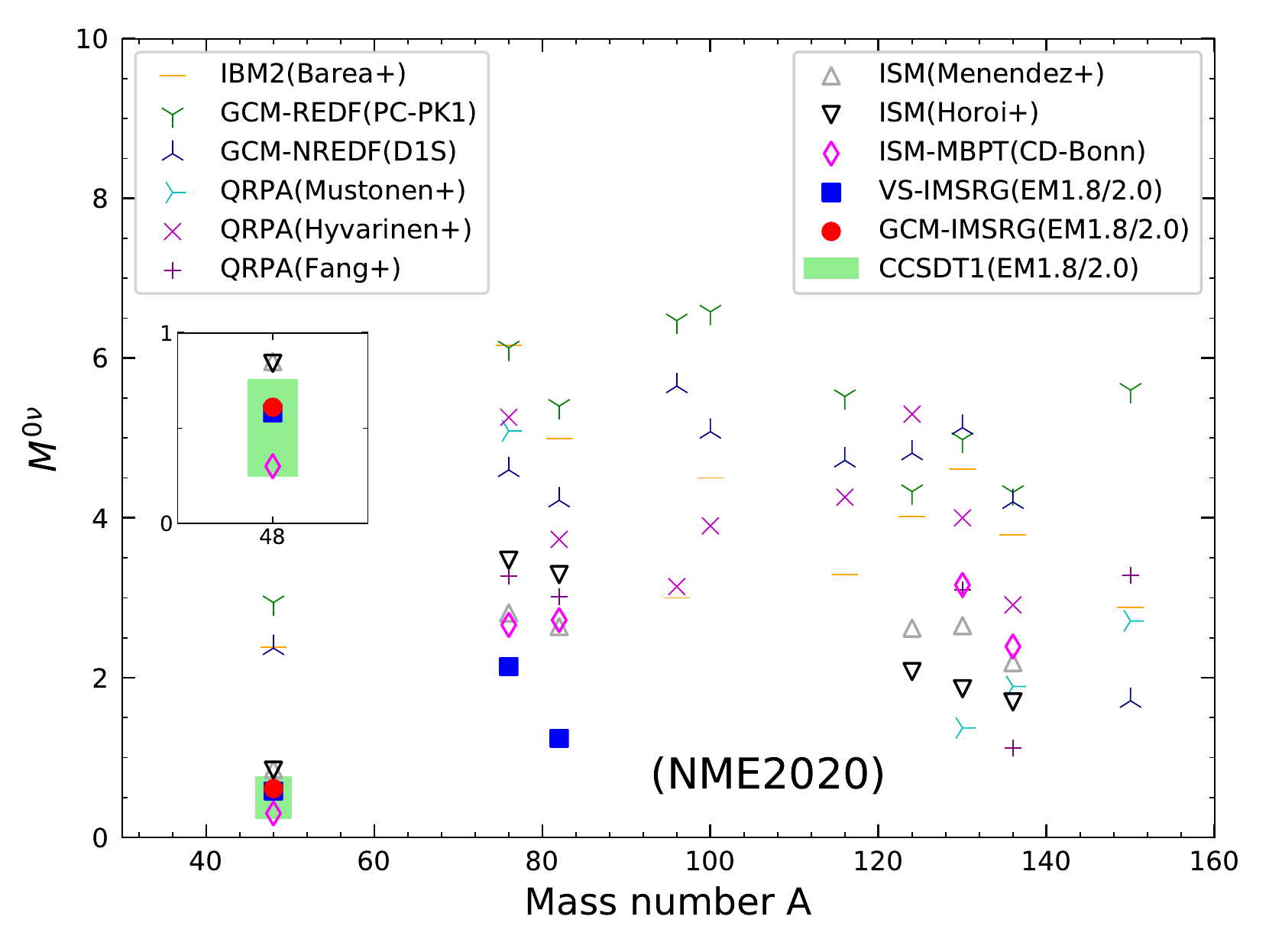} 
\caption{\label{fig:NMEs} (Color online) The NMEs of candidate $0\nu\beta\beta$ decay from the calculations of both phenomenological and {\em ab initio} nuclear models based on the mechanism of exchange light Majorana neutrinos. }
\end{figure}

Figure \ref{fig:NMEs} summarizes the  NMEs of $0\nu\beta\beta$ decay predicted by various nuclear models based on the standard light-Majorana neutirno-exchange mechanism. Compared to the values presented in a recent review paper \cite{Engel2017},  the updated NMEs from isospin-restored deformed quasi-particle random-phase-approximation (QRPA) calculation \cite{Fang2018}, the most recent ISM calculation using the Hamiltonian and decay operator constructed from many-body perturbation theory starting from CD-Bonn potential \cite{Coraggio2020} and the three {\em ab initio} calculations starting from the same chiral interaction \cite{Yao2020,Belley2020,Novario2020} are included. We note that the VS-IMSRG calculation also predicted NMEs for $^{76}$Ge and $^{82}$Se \cite{Belley2020}, which are to be confirmed with other {\em ab initio} calculations. It is shown that the NMEs by the {\em ab initio} calculations  are generally smaller than the predictions of phenomenological models, indicating that larger detectors may be required.

{\em ab initio} methods are essential for the calculation of $0\nu\beta\beta$ decay NMEs. However, there are many caveats while interpreting the NMEs from the reported {\em ab initio} calculations. These values are subject to change when the truncation error for nuclear interactions and transition operators in chiral EFT, and that in many-body methods are taken into account. For the transition operators, the contribution of a recently discovered  leading-order (LO) short-range operator \cite{Cirigliano2018PRL}, together with the next-to-next-to-leading order (N$^2$LO) operators \cite{Cirigliano2018PRC} that cannot be absorbed into momentum dependent form factors but could contribute up to $\mathcal{O}$(10\%) of the NME \cite{Pastrore2018}, is missing in the results presented in Fig. \ref{fig:NMEs}.  The LO short-range operator could either enhance or quench the NME by a amount that is comparable to its size \cite{Cirigliano2018PRL,Novario2020} depending on the unknown low-energy constant of this operator, which must be determined from future lattice QCD calculations \cite{Cirigliano2020}. Besides, the contribution of the two-body weak currents which induce three-nucleon interactions and appear at next-to-next-to-next-to-leading order (N$^3$LO) is not considered yet. It turns out that the two-body weak currents, together with many-body correlations, are important for the accurate calculation of nuclear single-beta decay rates \cite{Gysbers2019}. According to recent studies \cite{Menendez2011,Wang2018}, these two-body weak currents may also have a significant impact on the $0\nu\beta\beta$ decay NMEs. Contribution from other mechanisms, such as heavy-particle exchange, also needs to be examined. For the many-body truncation,  recent benchmark  studies against ``exact solution" from no-core shell-model calculations for very light nuclei \cite{Basili2020,Novario2020} provide a hint that the many-body truncation employed  in the reported {\em ab initio} calculations has a minor impact on the NMEs. This truncation error may become significant in the candidate decays of heavier nuclei and must be investigated.  Thus, lots of work still remains to be done to produce NMEs with fully quantified theoretical uncertainties.


 In summary, accurate NMEs for $0\nu\beta\beta$ decay of candidate nuclei are important for the design  and interpretation of future experiments.  Significant progress has been made in the modeling of these NMEs from first principles. The NME for $^{48}$Ca shows a good agreement among three different {\em ab initio} calculations starting from the same nuclear interaction constructed within the chiral EFT and the same decay operator. These studies open  the door to {\em ab initio} calculations of the matrix elements for the decays of heavier nuclei such as $^{76}$Ge, $^{130}$Te, and $^{136}$Xe. The ultimate goal is the computation of NMEs in many-body methods with controllable approximations using nuclear interactions and weak transition operators derived consistently from the chiral EFT with the feature of order-by-order convergence. We are expecting more progress towards this goal in the near future.

\paragraph{Conflict of interest}
The authors declare that they have no conflict of interest.

\paragraph{Acknowledgments.}
The author would like to thank J. Engel  and H. Hergert for a careful reading of the manuscript and constructive suggestions. Particular thanks go to B. Bally, C. L. Bai,  R.A.M. Basili, A. Belley,  G. Hagen, K. Hagino, J. D. Holt,  C. F. Jiao, J. Meng, S. Novario, T. Papenbrock, P. Ring, T. R. Rodriguez, S. R. Stroberg, L. S. Song, J. Vary, L. J. Wang and  R. Wirth for helpful discussions or collaborations at different stages. This work was supported in part by the U.S. Department of Energy, Office of Science, Office of Nuclear Physics under Awards No. DE-SC0017887, No. DE-FG02-97ER41019, No. DE-SC0015376 (the DBD Topical Theory Collaboration) and No. DE-SC0018083 (NUCLEI SciDAC-4 Collaboration).

\bibliography{references}

 \clearpage

\end{document}